\documentclass[preprint,12pt]{elsarticle}

\usepackage{amssymb}
\usepackage{amsmath}

\usepackage{subcaption}

\usepackage{hyperref}
\usepackage{float}
\usepackage{xcolor}
\newtheorem{theorem}{Theorem}

\newdefinition{rmk}{Remark}
\newdefinition{definition}{Definition}

\journal{Chaos, Solitons \& Fractals }

\begin{document}

\begin{frontmatter}



\title{A mathematical model of tumor growth using fractional derivatives}


\author[1,2]{Karen Escutia \corref{cor1}}
\ead{escutia_karen@ciencias.unam.mx}
\cortext[cor1]{Corresponding author}
\author[1,2]{Carlos Islas}
\ead{carlos.islas@ciencias.unam.mx}
\author[3]{Pablo Padilla}
\ead{pablo@mym.iimas.unam.mx}

\affiliation[1]{
    organization={ Universidad Nacional Autónoma de México},
    addressline={Circuito Exterior S/N, Ciudad Universitaria},
    city={Ciudad de México},
    postcode={04510},
    state={CDMX},
    country={México}
}
\affiliation[2]{
    organization={Universidad Autónoma de la Ciudad de México, Unidad Del Valle},
    addressline={San Lorenzo 290, Del Valle},
    city={Ciudad de México},
    postcode={03100},
    state={CDMX},
    country={México}
}

\affiliation[3]{
    organization={Instituto de Investigaciones en Matemáticas Aplicadas y en Sistemas, Departamento de Matemáticas y Mecánica, Universidad Nacional Autónoma de México},
    addressline={Circuito Exterior S/N, Ciudad Universitaria},
    city={Ciudad de México},
    postcode={04510},
    state={CDMX},
    country={México}
}

\begin{abstract}
In this work, we investigate a fractional-order tumor growth model aimed at capturing memory effects and nonlocal temporal dynamics inherent to tumor evolution. The model is formulated using Caputo fractional derivatives and incorporates key biological mechanisms related to tumor growth, vascular interaction, and cell clearance. To numerically solve the resulting fractional differential equations, a second-order fractional Runge--Kutta scheme is derived based on a truncated fractional Taylor expansion, providing an accurate and stable computational framework.

The proposed model is calibrated using experimental tumor volume data from five patients, and its performance is evaluated through the Root Mean Square Deviation (RMSD) between numerical simulations and experimental observations. The results show that, for all patients considered, the fractional-order model significantly improves the agreement with experimental data compared to the classical integer-order formulation. An optimal fractional order $\alpha<1$ is identified in each case, highlighting the relevance of memory effects in tumor growth dynamics and their patient-specific nature.

Further insight is obtained through phase--space and projection analyses, which reveal substantial geometric differences between integer-order and fractional-order dynamics. 

Although the present study is based on a limited number of patient datasets, the results demonstrate the potential of fractional-order modeling as a flexible and powerful framework for describing individualized tumor growth behavior. The proposed approach provides a solid basis for future extensions involving larger datasets, uncertainty quantification, and the incorporation of treatment effects and control strategies. \\


\end{abstract}







\end{frontmatter}




\section{Introduction}
\label{sec1}

Mathematical models of tumor growth, beyond their intrinsic interest, have become an important tool for the design of optimal treatment strategies.
These models aim to capture the essential biological mechanisms underlying cancer progression.

The term \textit{cancer} refers to a group of diseases characterized by the development of abnormal cells that divide, grow, and spread uncontrollably in any part of the body. These cells ignore the usual biological limits, acquiring the ability to proliferate continuously, invade nearby tissues, migrate to distant regions, and promote the formation of new blood vessels to obtain nutrients \cite{Gale2022, WHO_Cancer, Lipsick2020}.

Currently, cancer research has gained major relevance, as this disease represents one of the leading global health challenges and remains among the primary causes of mortality worldwide. According to recent data from the World Health Organization (WHO) \cite{WHO2024}, approximately 20 million new cases were diagnosed in 2022, with nearly 10 million deaths attributed to this group of diseases. Epidemiological studies further indicate that, over a lifetime, 1 in 5 individuals will develop some type of cancer, while cancer-specific mortality affects 1 in 9 men and 1 in 12 women.

In the quest for new strategies to achieve a deeper understanding of complex phenomena such as cancer, the mathematical modeling of biological processes has emerged as a fundamental tool for analyzing system dynamics. In particular, the scientific literature reports a broad spectrum of mathematical approaches that aim to better understand tumor growth dynamics. These studies range from basic phenomenological models \cite{martinlandrove2013, tejera2020, tejera2022, yousef2020} to highly detailed formulations that incorporate molecular mechanisms, including direct radiation-induced DNA damage \cite{Siam2018, Crooke2010}.

Moreover, tumor dynamics under different therapeutic interventions—such as immunotherapy, chemotherapy, and radiotherapy—has been the focus of continuous research \cite{Butner2022, SanchoAraiz2021, Anaya2021, Bekker2022, Zheng2025}. The interaction between treatment and tumor evolution can be modeled in several ways, incorporating both stochastic and deterministic processes, as well as more complex molecular mechanisms depending on the study’s objectives. In particular, multiple deterministic models analyze the effects of irradiation on tumor kinetics; these models typically represent tumor volume as composed of two constituents: proliferative cells and dead cells resulting from radiation exposure \cite{Zhong2014, Chvetsov2013, Huang2010, Lim2008}.

In this study, we focus on the model proposed by Watanabe et al.~\cite{Watanabe2016}, which describes the growth of solid tumors and their response to a stereoscopic radiosurgery process, a type of radiotherapy. This model considers that the tumor is composed of two types of cells: proliferative and non-proliferative. The key aspects of this approach are: (i) the tumor growth constant is assumed to be proportional to the radius of the vasculature, thus establishing a direct relationship between the supply of nutrients, oxygen, and the tumor expansion rate; and (ii) cells do not die instantly after receiving radiation but instead survive for several additional cell cycles. Furthermore, the model incorporates the radiation-induced death rate as the probability that a cell ceases its division, which was achieved by considering the survival described by the linear–quadratic model (LQ).

To represent the system dynamics, Watanabe et al.~\cite{Watanabe2016} use a set of ordinary differential equations that describe three stages of tumor growth: before irradiation, during irradiation, and after treatment. This formulation enabled advances in understanding tumor progression mechanisms and their response to radiosurgery therapy.

However, models formulated exclusively with ordinary differential equations present relevant limitations. In particular, they do not explicitly incorporate the influence of memory or the historical effects characteristic of many biological processes, which may play a crucial role in cancer progression and treatment effectiveness. Phenomena such as cell proliferation, angiogenesis, immune response, or cellular response to radiation depend not only on the current state of the system but also on its previous trajectory.

These considerations motivate the incorporation of mathematical tools capable of capturing such memory effects. In this context, fractional calculus offers a suitable formal framework to describe dynamics with historical dependence, nonlinear behavior, and complexity, making it a natural and promising extension of traditional models in mathematical oncology.

From a broader perspective, tumor growth can be naturally framed within the theory of nonlinear dynamical systems, where complex interactions, nonlocal effects, and memory play a fundamental role in shaping the evolution of biological processes. Within this setting, the combination of fractional-order operators with advanced numerical methods provides a powerful approach for analyzing such systems, allowing the incorporation of historical dependence while preserving a tractable computational structure. This perspective enables the investigation of tumor dynamics through tools commonly used in nonlinear science and numerical simulation, offering deeper insight into the mechanisms governing tumor evolution and treatment response.

Unlike classical calculus, which relies on integer-order derivatives and describes only instantaneous relationships between variables, fractional calculus extends this concept to non-integer (and even complex) orders, allowing the natural inclusion of memory and historical effects in the system. This characteristic makes it a powerful tool for describing phenomena with nonlocal dynamics and extended temporal dependence, aspects that are often present in biological, chemical, and medical systems~\cite{Dulf2015}.

In recent years, fractional calculus has gained significant relevance in the modeling of biomedical processes, demonstrating a greater ability to capture long-term memory, hysteresis, and anomalous responses compared with integer-order models~\cite{Dulf2019, Yao2021, Deepika2023, Dulf2012}. Thanks to its flexible nature, fractional models allow a more faithful reproduction of the real dynamic trajectories of living systems, in which processes such as transport, diffusion, or cell growth do not necessarily follow classical laws.

In particular, in the study of tumor growth, the incorporation of fractional dynamics has shown highly promising results. For example, Alinei-Poiana et al.~\cite{AlineiPoiana2023} analyzed four classical tumor-growth models —exponential, logistic, Gompertz, and Bertalanffy–Pütter— demonstrating that the introduction of fractional operators significantly improves the predictive accuracy of tumor growth, validated through experimental data obtained from mice.

Additionally, studies integrating exponential-type fractional memories and research exploring the presence of chaos in tumor growth models have been reported, such as the work by Krunal Kachia et al.~\cite{Kachia2020}.

Furthermore, authors such as Seda İğret Araz~\cite{IgretAraz2021} and İrem Akbulut Arık et al.~\cite{Arik2022} have extended the model of Watanabe et al.~\cite{Watanabe2016} by incorporating different fractional operators, which has allowed a more realistic description of dynamic transitions and memory effects in tumor evolution.

These advances show that fractional models not only broaden the theoretical framework of classical differential calculus, but also provide a more complete, coherent, and physiologically grounded perspective for understanding complex biological phenomena, particularly those characterized by nonlocal processes and history dependence.

The present work aims to combine the tumor growth model before and after a radiosurgery process, proposed by Watanabe et al.~\cite{Watanabe2016}, with the use of fractional differential operators. The models will be solved numerically through the implementation of a second-order fractional Runge–Kutta method (FORK), and the results obtained will be compared between the classical and fractional approaches using data from five patients with metastatic brain cancer treated with stereotactic radiosurgery (\emph{Gamma Knife}). The comparison will be performed using the calculation of the \emph{Root Mean Square Deviation} (RMSD) as a quantitative measure of the fit.

This article is structured as follows: after this introduction, Section \textbf{Mathematical Preliminaries}~\ref{sec:preliminares} presents the necessary definitions and foundations of fractional calculus. Next, in Section \textbf{Formulation of the fractional differential equation model}~\ref{sec:formulation} we describe the considerations for the formulation of systems of fractional differential equations for cancer growth model and the section  \textbf{Fractional Runge-Kutta Method}~\ref{sec:runge}, the development of the FORK method applied to systems of fractional differential equations is described. Subsequently, in Section \textbf{Results and Discussion}~\ref{sec:resultados}, the main findings of the study are analyzed, and finally, the work concludes with Section \textbf{Conclusions}~\ref{sec:concl}, where the contributions and future perspectives are summarized.

\section{Mathematical Preliminaries} \label{sec:preliminares}

Within the framework of fractional calculus, there exist multiple definitions of \emph{integral operators}~\cite{Monje2010}.  
One of the most widely used definitions is that of \emph{Riemann–Liouville}, which establishes the fractional-order integral with $\Re(\alpha) > 0$ as a natural consequence of Cauchy's formula for iterated integrals, written as:

\begin{equation}
_aI^n_t f(t) = \frac{1}{(n-1)!} \int_{a}^{t} (t-\tau)^{n-1} f(\tau) \, d\tau 
\end{equation}

where $I$ denotes the integral operator and $n \in \mathbb{N}$ represents the order of integration.  
By introducing the Gamma function into the previous equation, the notion of integration order can be extended from the set of natural numbers to the set of positive real numbers. Thus, the fractional integral is defined in the sense of \emph{Riemann–Liouville}.

\begin{definition}[Riemann--Liouville Fractional Integral \cite{Monje2010}] 
Let $f \in L^1[a,t]$ and $\alpha > 0$.  
The fractional integral of order $\alpha$ of $f$ in the sense of Riemann--Liouville is defined as:
\begin{equation} \label{eqn:integral_riemman}
(^{RL}_{a}I_{t}^\alpha f)(t) = \frac{1}{\Gamma(\alpha)} \int_{a}^{t} (t-\tau)^{\alpha - 1} f(\tau) \, d\tau 
\end{equation}
where $\alpha > 0$ is the new integration order and
\begin{equation} \label{eqn:gamma}
\Gamma(\alpha) = \int_{0}^{\infty} t^{\alpha-1} e^{-t} \, dt
\end{equation}
is the Gamma function, which generalizes the factorial function.

\end{definition}

Using the previous definition, we can define the fractional derivative in the sense of Riemann--Liouville.

\begin{definition}[Riemann--Liouville Fractional Derivative \cite{Monje2010}]
Let $n \in \mathbb{N}$ such that $n-1 < \alpha \leq n$.  
The fractional derivative of order $\alpha$ of $f$ in the sense of Riemann--Liouville is defined as:
\begin{equation}\label{eqn:derivada_riemman}
(^{RL}_{a}D_{t}^\alpha f)(t) =
\frac{1}{\Gamma(n-\alpha)}
\frac{d^n}{dt^n}
\int_{a}^{t} (t-\tau)^{n-\alpha-1} f(\tau) \, d\tau 
\end{equation}
\end{definition}

An alternative and widely used form is the fractional derivative in the sense of Caputo. 

\begin{definition}[Caputo Fractional Derivative \cite{Chakraverty2022}]
Let $n \in \mathbb{N}$ such that $n-1 < \alpha \leq n$.  
The fractional derivative of order $\alpha$ of $f$ in the sense of Caputo is defined as:
\begin{equation}\label{eqn:derivada_caputo}
(^{C}_{a}D_{t}^\alpha f)(t) =
\frac{1}{\Gamma(n-\alpha)}
\int_{a}^{t} (t-\tau)^{n-\alpha-1} \frac{d^n f(\tau)}{dt^n} \, d\tau 
\end{equation}
\end{definition}

The Caputo fractional derivative has a significant advantage over other definitions of fractional derivatives, such as Riemann--Liouville, due to the way it incorporates initial conditions. In the Caputo sense, these conditions are expressed in terms of ordinary integer-order derivatives, which maintains consistency with the classical formulation of initial value problems. 

This feature facilitates both the physical interpretation and the direct application of fractional models to real systems, especially those that can be described by empirical laws or memory-dependent processes. Moreover, using the Caputo definition allows a more natural transition between classical models and their fractional extensions, preserving the physical meaning of initial conditions.

With this in mind, we considered that tumor growth models are described not only through fractional derivatives but also through \textit{fractional differential equations}, which may or may not be linear, generally taking the form:

\begin{equation}
\label{eqn: ecuacion_general}
{}^{C}_a D_t^\alpha f(t) = \phi(t, f(t)), \quad f(0) = f_0, \quad \alpha \in (0,1]
\end{equation}

Due to the complexity of obtaining analytical solutions for such models, various numerical approaches have been developed to solve them efficiently. One of the best-known methods is the Adams predictor–corrector method \cite{Wang2014}, which provides accurate results but with high computational cost.

In this work, we propose an alternative based on a two-stage fractional Runge–Kutta scheme \cite{Arshad2020, Ghoreishi2023}, which allows obtaining numerical solutions with good accuracy and lower computational demand.

To develop the Runge–Kutta method, the following results will be required.

\begin{theorem}[Generalized Mean Value Theorem \cite{Ghoreishi2023}] 
Suppose \( f(x) \in C[a,b] \) and \( ^C _aD_t^{\alpha} f(t) \in C(a,b] \) for \( 0 < \alpha \leq 1 \). Then,
\begin{equation} \label{eqn:valormedio_caputo}
f(t) = f(a) + \frac{1}{\Gamma(\alpha + 1)} ({}^{C}D_a^\alpha f)(\xi) (t - a)^\alpha, \quad \text{with } a \leq \xi \leq t, \; t \in (a,b]
\end{equation}
\end{theorem}

\begin{theorem}[Generalized Taylor Formula in the Caputo Sense \cite{Ghoreishi2023}]
Suppose that ${}_a^{C}D_t^{k\alpha} f(t) \in C(a,b]$ for $k = 0,1,\dots,n+1$, where
$0 < \alpha \leq 1$. Then,
\begin{equation}\label{eqn:taylor_generalizado}
f(t)
=
\sum_{i=0}^{n}
\frac{(t-a)^{i\alpha}}{\Gamma(i\alpha+1)}
\left({}_a^{C}D_t^{i\alpha} f\right)(a)
+
\frac{(t-a)^{(n+1)\alpha}}{\Gamma((n+1)\alpha+1)}
\left({}_a^{C}D_t^{(n+1)\alpha} f\right)(\xi),
\end{equation}
where $a < \xi < t$ and $t \in (a,b]$.

Moreover, the fractional Taylor expansion of a composite function
$\phi(t,u(t))$ about $t=t_0\in(a,b]$ can be formally written as
\begin{equation}\label{eq:frac_taylor_phi_t0}
\begin{aligned}
\phi(t,u(t)) ={}&
\phi(t_0,u(t_0))
+ \frac{(t-t_0)^{\alpha}}{\Gamma(\alpha+1)}
\,{}_a^{C}D_t^{\alpha}\phi(t_0,u(t_0))  \\
&+ (u(t)-u(t_0))\,\partial_u \phi(t_0,u(t_0))  \\
&+ \frac{(t-t_0)^{2\alpha}}{\Gamma(2\alpha+1)}
\,{}_a^{C}D_t^{2\alpha}\phi(t_0,u(t_0)) \\
&+ \frac{(u(t)-u(t_0))^2}{2!}\,\partial_u^2 \phi(t_0,u(t_0)) \\
&+ \frac{(t-t_0)^{\alpha}(u(t)-u(t_0))}{\Gamma(\alpha+1)}
\,{}_a^{C}D_t^{\alpha}\partial_u \phi(t_0,u(t_0)) \\
&+ \frac{(t-t_0)^{3\alpha}}{\Gamma(3\alpha+1)}
\,{}_a^{C}D_t^{3\alpha}\phi(t_0,u(t_0))
+ \cdots 
\end{aligned}
\end{equation}

The operator ${}_a^{C}D_t^{\alpha}$ denotes the Caputo fractional derivative
of order $\alpha$ with respect to the time variable $t$, while
$\partial_u$ denotes the classical (integer-order) partial derivative
with respect to the state variable $u$.
\end{theorem}

\begin{theorem} \label{eqn.Leibniz}
Let $0 < \alpha < 1$, and suppose $f$ and $g$ are analytic in the interval $(a-h,\,a+h)$. 
Then, the Leibniz formula for the Caputo derivative is

\begin{equation}
{}_{a}^C D_t^{\alpha}\!\left( \Phi(t)f(t)\right)
=
\sum_{k=0}^{\infty}  \binom{\alpha}{k} \Phi^k (t) 
{}_{a}^{RL} D_t^{\,\alpha-k} f(t)
\;-\;
\Phi(a)f(a)\,\frac{(t-a)^{-\alpha}}{\Gamma(1-\alpha)} .
\end{equation}
where 
\[
\binom{\alpha}{k} \;=\;
\frac{\Gamma(\alpha+1)}{\Gamma(k+1)\,\Gamma(\alpha-k+1)},
\qquad \alpha\in\mathbb{R},\; k\in\mathbb{N}.
\]
\end{theorem}

\section{Formulation of the fractional differential equation model} \label{sec:formulation}

The goal of this study is to evaluate tumor growth dynamics and its response to radiation treatment through a system of fractional differential equations, inspired by the work of Watanabe \textit{et al.} \cite{Watanabe2016}. The proposed model describes the temporal evolution of tumor volume before, during, and after a radiosurgery process, incorporating fractional dynamics as a tool to capture biological memory effects and the temporal nonlocality of the system.

The fractional model considers the presence of two cellular types: \textit{proliferative cells}, responsible for the active growth of the tumor, and \textit{non–proliferative cells}, which emerge as a result of radiation treatment. The slowdown in tumor growth as its volume increases is modeled by assuming that the growth rate depends on the relationship between vascular and tumor volumes, and that vasculature develops at a slower rate than the tumor itself.

This proposal can be represented by a set of three fractional differential equations with three variables: the volume of proliferative tumor tissue $V_T(t)$, the volume of non–proliferative cells $V_{ND}(t)$, and the tumor growth rate $\lambda(t)$, which is not constant but time–dependent, and four constants: a radiobiological parameter $\alpha$, the initial growth rate $\lambda(0)$, the vascular growth–delay factor $\theta$, and the cellular destruction rate $\eta_{cl}$.

The natural growth of the tumor can be represented by the following system of fractional differential equations, where an exponential-type tumor growth is assumed that depends only on the growth rate which, in turn, decreases proportionally to the vascular growth–delay factor:

\begin{align}
{}^C D^\gamma V_T (t) &= \lambda(t) V_T  \\
{}^C D^\gamma \lambda(t) &= -\theta \lambda(0) \lambda 
\end{align}

Under the influence of a radiation dose $D$, dividing cells may continue to proliferate with a tumor growth rate $\lambda(t)$ multiplied by the cellular proliferation probability $p(D)$, or they may transition to a non-dividing state at a rate $g(D)$. Non-proliferative cells are eventually removed at a cellular destruction rate $\eta_{cl}$. The radiation-induced cellular destruction rate is determined by the probability that a cell stops dividing at the end of the cell cycle; this probability is directly related to the surviving fraction given by the linear–quadratic  model (LQ). Moreover, the response to radiation is formulated by considering that cells do not die instantaneously, but may survive for several cell cycles before losing their proliferative capacity. This aspect is crucial to justify the incorporation of fractional dynamics in the model, since the future state of each cell depends on its prior history: the system possesses \emph{memory}. In this regard, fractional calculus provides an appropriate mathematical framework to describe such temporal dependence and to more accurately capture the biological evolution of the tumor under radiotherapeutic treatment.

Assuming the dose $D$ is applied at time $t_R$, there is a period during which the radiation takes effect, denoted $\tau_{\text{rad}}$, i.e., for $t_R \leq t < t_R + \tau_{\text{rad}}$:
\begin{align}
{}^C D^\gamma V_T &= \lambda(t)\, p(D)\, V_T - g(D)\, V_T,  \\
{}^C D^\gamma V_{ND} &= g(D)\, V_T - \eta_{\text{cl}}\, V_{ND},  \\
{}^C D^\gamma \lambda &= -\theta \lambda(0)\, \lambda .
\end{align}

Finally, for times when the radiation no longer affects the cells, the tumor volume no longer undergoes active reduction and the non-proliferative cells begin to decline naturally; that is, for $t > t_R + \tau_{\text{rad}}$, $V_T$, $V_{ND}$ and $\lambda$ are solutions of the following system:
\begin{align}
{}^C D^\gamma V_T(t) &= \lambda(t)\, V_T,  \\
{}^C D^\gamma V_{ND} (t) &= -\eta_{\text{cl}}\, V_{ND},  \\
{}^C D^\gamma \lambda (t) &= -\theta \lambda(0)\, \lambda .
\end{align}

The two radiation–dose–dependent parameters—the probability of cellular death $p(D)$ and the transition rate from proliferative to non-proliferative tumor cells $g(D)$—according to Watanabe et al. \cite{Watanabe2016}, are given by:

\begin{align}
p(D) &= 1 - \frac{T^*}{3 T_m}\, \chi(D),  \\
g(D) &= \frac{\chi(D)}{3 T_m},
\end{align}

where the function $\chi(D)$ comes from the linear–quadratic (LQ) model:

\begin{equation}
\chi(D) = \alpha_{LQ} D \left(1 + \frac{D}{\alpha_{LQ}/\beta}\right).
\end{equation}

The times used satisfy $T^* \approx T_{cc}$ for practical purposes, where $T_{cc}$ represents the characteristic time or the duration of a cell cycle, and $T_m$ is the observation time.

To assess the effectiveness of the fractional models, data from five patients treated with radiosurgery were used. The most relevant experimentally measured information regarding the evolution of their tumor development is presented in Table \ref{tab:uno}.

\begin{table}[H]
\centering
\begin{tabular}{|c|c|ccccc|}
\hline
\textbf{Parameters} & \textbf{Unit} & \textbf{Case 1} & \textbf{Case 2} & \textbf{Case 3} & \textbf{Case 4} & \textbf{Case 5} \\ 
\hline
\textbf{Primary cancer} &  & NSCL & RCC & Melanoma & RCC & RCC \\
\hline
Inicial tumor volume & $cm^3$ & 0.126 & 0.178 & 0.101 & 1.69 & 0.0065 \\
Prescription dose & Gy & 22 & 18 & 20 & 13 & 16 \\
Tumor volume at GKSRS & $cm^3$ & 0.258 & 2.626 & 0.871 & 3.86 & 5.5 \\
Tumor volume at end & $cm^3$& 0.019 & 7.759 & 20.560 & 3.016 & 1.236 \\
Total monitor duration & days & 137 & 419 & 80 & 471 & 680 \\
Day of GKSRS & days & 34 & 103 & 29 & 239 & 117 \\
\hline
\end{tabular}
\caption{Treatment-related parameters.\\NSCL- non small cell lung cancer, RCC renal cell carcinoma }
\label{tab:uno}
\end{table}

The four model parameters, namely the fractional order $\alpha$, the initial growth rate $\lambda(0)$, the angiogenic parameter $\theta$, and the clearance parameter $\eta_{cl}$, were estimated through a fitting procedure based on the total tumor volume, defined as the sum of $V_T$ and $V_{ND}$. 

The parameter $\eta_{cl}$ is related to the cell clearance time $t_{cl}$ through the relation $t_{cl} = 0.693/\eta_{cl}$. Similarly, the tumor volume doubling time $t_d$ is associated with the effective growth rate $\lambda$, which depends on both the tumor volume and the vascular structure, and is given by $t_d = 0.693/\lambda$.

The parameter estimation was carried out by implementing the \textit{Differential Evolution} algorithm in the Python programming language.  
This global optimization method seeks to minimize an objective function, which in this case corresponds to the Root Mean Square Deviation (RMSD) between the experimental data and the solutions obtained from the fractional model.  
Mathematically, the RMSD is defined as

\begin{equation}
\mathrm{RMSD} = \sqrt{\frac{1}{N} \sum_{i=1}^{N} \left( y_i^{\text{exp}} - y_i^{\text{mod}} \right)^2 }
\end{equation}

where \( y_i^{\text{exp}} \) denotes the experimental values, \( y_i^{\text{mod}} \) the values computed from the model, and \( N \) the total number of observations.

Since the system describes a biological process, all model parameters were assumed to be positive to preserve the physical and biological consistency of the simulations.

\section{Second-Order Fractional Runge--Kutta Method}\label{sec:runge}

Numerical solutions of the proposed fractional-order model are obtained using a Runge–Kutta–based scheme adapted for fractional differential equations is an iterative method that delivers solutions with high precision and efficiency. In recent years, particular emphasis has been placed on extending this model to solve fractional differential equations \cite{Arshad2020, Ghoreishi2023}.

Below, we present the derivation of the two-stage Runge–Kutta iterative method for solving fractional differential equations that are not necessarily linear.

To derive the fractional Runge–Kutta method, it is necessary to consider the fractional time derivative in the Caputo sense, which allows us to use classical initial conditions. Consider the following problem:

\begin{equation}
\label{eqn: ecuaciones_general}
{}^{C}D_t^\alpha f(t) = \phi(t, f(t)), \qquad f(0) = f_0, \qquad \alpha \in (0,1]
\end{equation}

Let $[0,b]$ be the interval over which we seek the solution to problem (\ref{eqn: ecuaciones_general}). The interval is subdivided into $r$ subintervals $[t_j, t_{j+1}]$ of uniform step size $h = \frac{b}{r}$ using nodes $t_j = jh$, with $j = 0,1,2,\ldots,r$.

Now, if we consider the generalized Taylor formula defined in \ref{eqn:taylor_generalizado}:

\[
f(t) = \frac{({}^{C}_aD_t^{(n+1)\alpha} f)(\xi)}{\Gamma((n+1)\alpha + 1)} (t - a)^{(n+1)\alpha} + \sum_{i=0}^{n} \frac{(t - a)^{i\alpha}}{\Gamma(i\alpha + 1)} ({}^{C}_aD_t^{i\alpha} f)(a)
\]

moreover, for each time increment $h$, we have:

\[
f(t + h) = \frac{({}^{C}_aD_t^{(n+1)\alpha} f)(\xi)}{\Gamma((n+1)\alpha + 1)} ((t + h) - t)^{(n+1)\alpha} + \sum_{i=0}^{n} \frac{((t + h) -t)^{i\alpha}}{\Gamma(i\alpha + 1)} ({}^{C}_aD_t^{i\alpha} f)(t) 
\]
\[
=  \frac{({}^{C}_aD_t^{(n+1)\alpha} f)(\xi)}{\Gamma((n+1)\alpha + 1)} h^{(n+1)\alpha} + \sum_{i=0}^{n} \frac{h^{i\alpha}}{\Gamma(i\alpha + 1)} ({}^{C}_aD_t^{i\alpha} f)(t) 
\]

Assuming $f(t)$, ${}^{C}_aD_t^\alpha f(t)$ and ${}^{C}_a D_t^{2\alpha} f(t)$ are continuous functions on $[0,b]$, we obtain:

\begin{equation} \label{eqn: expansion}
f(t + h) = f(t) + \frac{h^\alpha}{\Gamma(\alpha + 1)} {}^{C}_aD_t^\alpha f(t) 
+ \frac{h^{2\alpha}}{\Gamma(2\alpha + 1)} {}^{C}_aD_t^{2\alpha} f(t)  + \cdots
\end{equation}

If we take the derivative of order $\alpha$ and $2\alpha$ of the differential equation of our problem \ref{eqn: ecuaciones_general}, we obtain:

\[
{}^{C}_aD_t^{\alpha} f(t) =  \phi(t, f(t)) 
\]

\[
{}^{C}_aD_t^{2\alpha} f(t) = {}^{C}_aD_t^\alpha \phi(t, f(t))
\]

To obtain the expression for the derivative of order $2\alpha$, we used the chain rule to obtain the following formula:

\begin{equation} \label{eqn: 2alpha}
{}^{C}_aD_t^{2\alpha} f(t)
=
{}^{C}_aD_t^{\alpha} \phi(t, f(t))
+
\phi(t,f(t))\,\partial_f \phi(t,f(t))
\end{equation}
 Substituting \ref{eqn: 2alpha} into the previous expansion \ref{eqn: expansion} yields the reorganized form:

\begin{equation} \nonumber
f(t + h) = f(t) + \frac{h^\alpha}{\Gamma(\alpha + 1)} {}^{C}_aD_t^\alpha f(t) 
+ \frac{h^{2\alpha}}{\Gamma(2\alpha + 1)} {}^{C}_aD_t^{2\alpha} f(t) + \cdots 
\end{equation}
\begin{equation}
= f(t) + \frac{h^\alpha}{\Gamma(\alpha + 1)} {}^{C}D_t^\alpha f(t) 
+ \frac{h^{2\alpha}}{\Gamma(2\alpha + 1)} \left({}^{C}D_t^\alpha \phi(t, f(t)) + \phi(t,f(t)) \, \partial_f \phi(t,f(t))\right) + \cdots
\end{equation}

Assuming a sufficiently small step size, we neglect the higher-order terms involving the Caputo fractional derivative ${}^{C}D_t^{3\alpha}$, we obtain the following formulas

\begin{equation} \nonumber
f(t + h) = f(t) + \frac{h^\alpha}{\Gamma(\alpha + 1)} {}^{C}_aD_t^\alpha f(t) 
+ \frac{h^{2\alpha}}{\Gamma(2\alpha + 1)} \left({}^{C}_aD_t^\alpha \phi(t, f(t)) + \phi(t,f(t)) \, \partial_f\phi(t,f(t))\right) 
\end{equation}

\begin{equation} \nonumber
\begin{aligned}
= f(t) + \frac{h^\alpha}{2\Gamma(\alpha + 1)} {}^{C}_aD_t^\alpha f(t)  
\end{aligned}
\end{equation}

\begin{equation} \nonumber
+ \frac{h^\alpha}{2\Gamma(\alpha + 1)} \left({}^{C}_aD_t^\alpha f(t) 
+ \frac{2 h^{\alpha} \Gamma(\alpha + 1) }{\Gamma(2\alpha + 1)} \left({}^{C}_aD_t^\alpha \phi(t, f(t)) + \phi(t,f(t)) \, \partial_f \phi(t,f(t))\right) \right)
\end{equation}

\begin{equation} \nonumber
\begin{aligned}
= f(t) + \frac{h^\alpha}{2\Gamma(\alpha + 1)} \phi(t,f(t)) 
\end{aligned}
\end{equation}

\begin{equation} \label{eqn:aproximate}
+ \frac{h^\alpha}{2\Gamma(\alpha + 1)} \left(\phi(t,f(t))
+ \frac{2 h^{\alpha} \Gamma(\alpha + 1) }{\Gamma(2\alpha + 1)} \left({}^{C}_aD_t^\alpha \phi(t, f(t)) + \phi(t,f(t)) \, \partial_f \phi(t,f(t))\right) \right)
\end{equation}

Motivated by the fractional Taylor expansion of $\phi$ around $(t,f(t))$ show in \ref{eq:frac_taylor_phi_t0}
we introduce the fractional increments
\begin{equation}\label{eq:increments}
\Delta t =
\left(
\frac{2 h^{\alpha}\Gamma(\alpha+1)^2}{\Gamma(2\alpha+1)}
\right)^{1/\alpha},
\qquad
\Delta f =
\left(
\frac{2 h^{\alpha}\Gamma(\alpha+1)}{\Gamma(2\alpha+1)}
\right)\phi(t,f(t))
\end{equation}

Then, the following approximation holds:
\begin{equation}\label{eq:phi_shift}
\phi(t+\Delta t,\,f(t)+\Delta f)
\approx
\phi(t,f(t))
+ 
\frac{2 h^{\alpha}\Gamma(\alpha+1)}{\Gamma(2\alpha+1)}
\left[
{}^{C}_aD_t^\alpha \phi(t,f(t))
+
\phi(t,f(t))\,\partial_f \phi(t,f(t))
\right]
\end{equation}

Substituting \eqref{eq:phi_shift} into \eqref{eqn:aproximate}, we obtain the
fractional two-stage approximation
\begin{equation}\label{eq:final_scheme}
f(t+h)
\approx
f(t)
+
\frac{h^\alpha}{2\Gamma(\alpha+1)}
\left[
\phi(t,f(t))
+
\phi(t+\Delta t,\,f(t)+\Delta f)
\right]
\end{equation}

\begin{equation}
 = f(t) + \frac{h^\alpha}{2\Gamma(\alpha + 1)} \phi(t, f(t)) 
+ \frac{h^\alpha}{2\Gamma(\alpha + 1)} \phi\left(t + \left(
\frac{2 h^{\alpha}\Gamma(\alpha+1)^2}{\Gamma(2\alpha+1)}
\right)^{1/\alpha}, f(t) + \frac{2 h^{\alpha}\Gamma(\alpha+1)}{\Gamma(2\alpha+1)} \phi(t,f) \right) 
\end{equation}

In view of the previous expression, the two–stage fractional Runge–Kutta method is obtained:

\begin{equation}
f_{n+1} = f_n + \frac{h^\alpha}{2\Gamma(\alpha + 1)} \big( K_1 + K_2 \big),
\end{equation}
where
\[
K_1 = \phi(t_n, f_n),
\]
\[
K_2 = \phi\left(t_n + \left(
\frac{2 h^{\alpha}\Gamma(\alpha+1)^2}{\Gamma(2\alpha+1)}
\right)^{1/\alpha},  \; f_n + \frac{2 h^{\alpha}\Gamma(\alpha+1)}{\Gamma(2\alpha+1)} K_1 \right) 
\]

If $\alpha = 1$, this scheme reduces to the classical second–order Runge–Kutta method.

\section{Results and Discussion}\label{sec:resultados}
Five model parameters, namely the fractional order $\alpha$, the linear–quadratic radiobiological parameter $\alpha_{LQ}$, the tumor doubling time $t_d$, the angiogenic parameter $\theta$, and the cell clearance time $t_{cl}$, were optimized to achieve the best fit to the experimental data. The remaining radiobiological parameters were fixed as $\alpha_{LQ}/\beta = 10$~Gy, $\tau_{rad} = 8$~days, $T_{cc} = 1$~day, and $T_m = 10$~days, following the values reported in~\cite{Watanabe2016}.

The estimated model parameters for the five patients are summarized in Table~\ref{tab:dos}.

\begin{table}[H]
\centering
\begin{tabular}{|c|c|ccccc|}
\hline
\textbf{Parameters} & \textbf{Unit} & \textbf{Case 1} & \textbf{Case 2} & \textbf{Case 3} & \textbf{Case 4} & \textbf{Case 5} \\ 
\hline
\textbf{Primary cancer} &  & NSCL & RCC & Melanoma & RCC & RCC \\
\hline
$\alpha_{LQ}$ & 1/Gy & 0.14 & 0.025 & 0.011 & 0.19 & 0.18 \\
$\theta$ & & 0.62 & 0.31 & 0.16 & 0.01 & 0.12 \\
$t_{cl}$ & days & 13.0 & 28.0 & 15.0 & 250.0 & 38.0 \\
$t_d$ & days & 29.0 & 8.0 & 9.0 & 230.0 & 7.0 \\
$V_{T0}$ & cm$^3$ & 0.126 & 0.178 & 0.101 & 1.69 & 0.0065 \\
$t_R$ & days & 34.0 & 103.0 & 29.0 & 239.0 & 117.0 \\
$D$ & Gy & 22.0 & 18.0 & 20.0 & 13.0 & 16.0 \\
\hline
\end{tabular}
\caption{Treatment-related parameters.\\NSCL- non small cell lung cancer, RCC renal cell carcinoma }
\label{tab:dos}
\end{table}

\begin{figure}[H]
    \centering
    
    \begin{subfigure}{0.48\textwidth}
        \includegraphics[width=\linewidth]{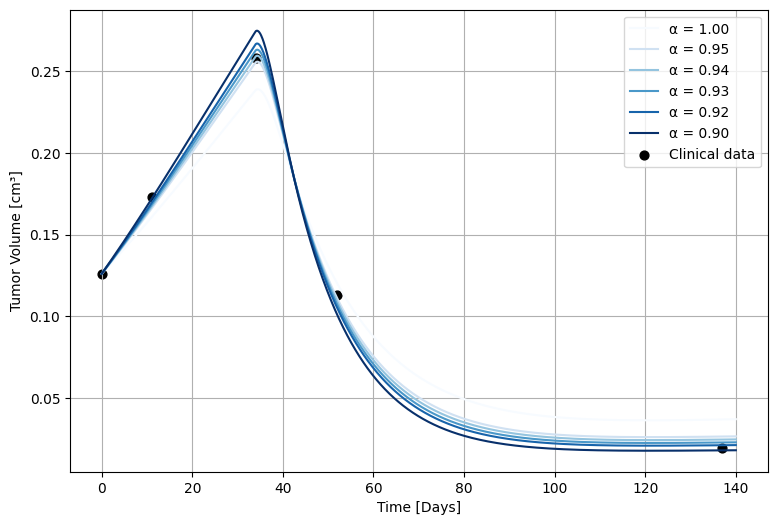}
        \caption{}
    \end{subfigure}
    \begin{subfigure}{0.48\textwidth}
        \includegraphics[width=\linewidth]{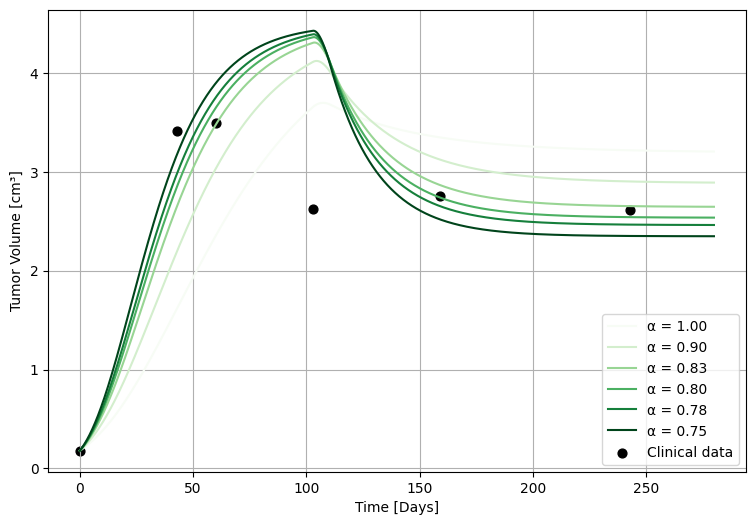}
        \caption{}
    \end{subfigure}
    \begin{subfigure}{0.48\textwidth}
        \includegraphics[width=\linewidth]{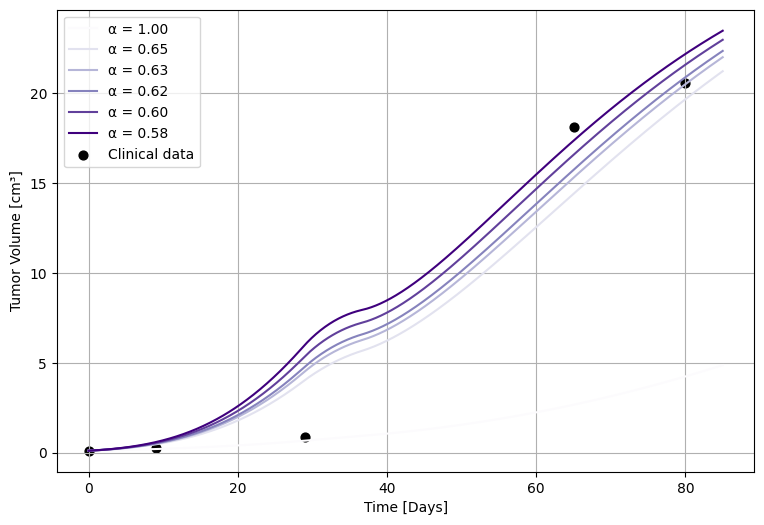}
        \caption{}
    \end{subfigure}
    \begin{subfigure}{0.48\textwidth}
        \includegraphics[width=\linewidth]{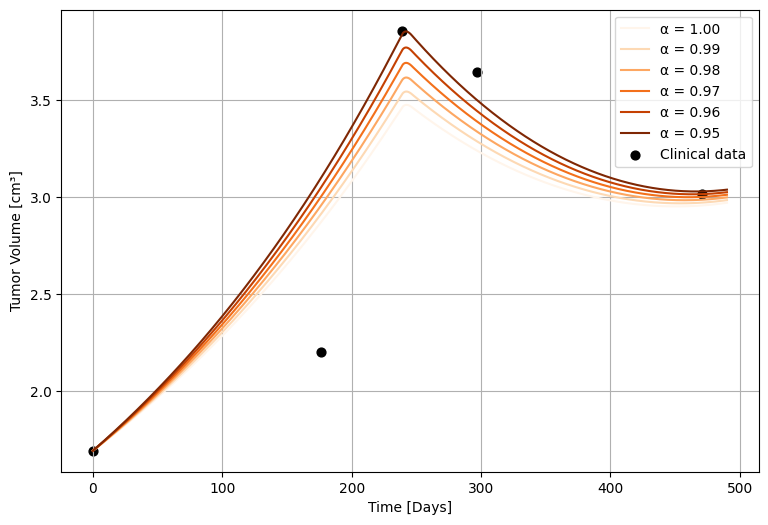}
        \caption{}
    \end{subfigure}
    \begin{subfigure}{0.48\textwidth}
        \includegraphics[width=\linewidth]{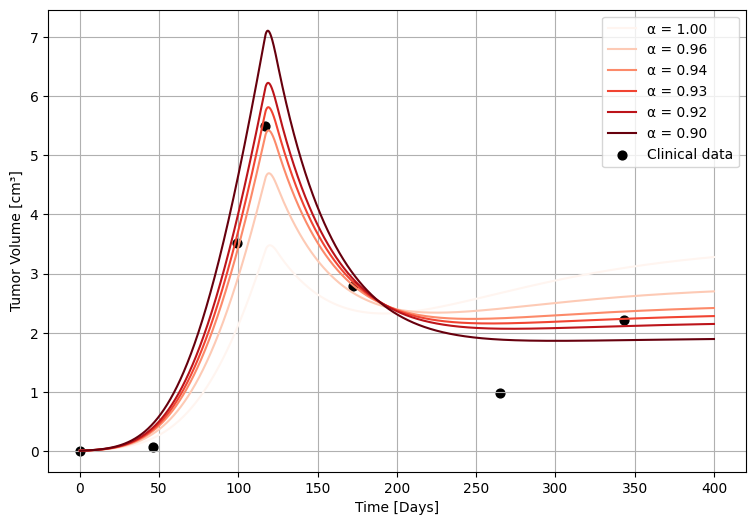}
        \caption{}
    \end{subfigure}

    \caption{Tumor volume evolution for different fractional orders in the case of the: (a) patient one; (b) patient two; (c) patient three; (d) patient four; (e) patient five}
    \label{fig:resultados}
\end{figure}

The simulation results are shown in Figure~\ref{fig:resultados}. These plots illustrate the temporal evolution of the tumor volume obtained from the numerical solution of the proposed model using the second-order fractional Runge--Kutta method described in the previous section, which incorporates fractional dynamics. Six different approximations corresponding to distinct integration orders are presented, including the classical integer-order case ($\alpha = 1$). Figure~\ref{fig:resultados} reveals that the fractional order $\alpha$ has a significant impact on tumor growth dynamics, leading to markedly different temporal evolutions of the tumor volume.

\begin{figure}[H]
    \centering
    
    \begin{subfigure}{0.48\textwidth}
        \includegraphics[width=\linewidth]{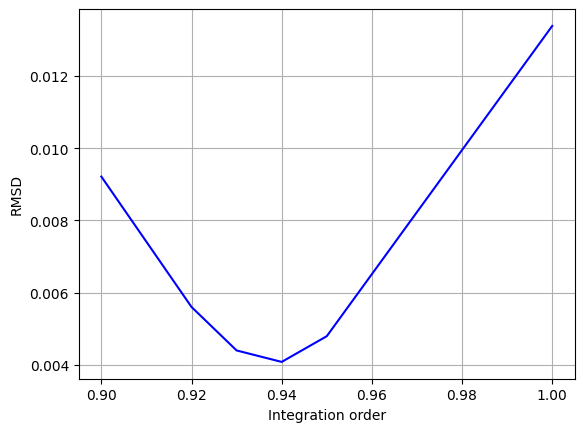}
        \caption{}
    \end{subfigure}
    \begin{subfigure}{0.48\textwidth}
        \includegraphics[width=\linewidth]{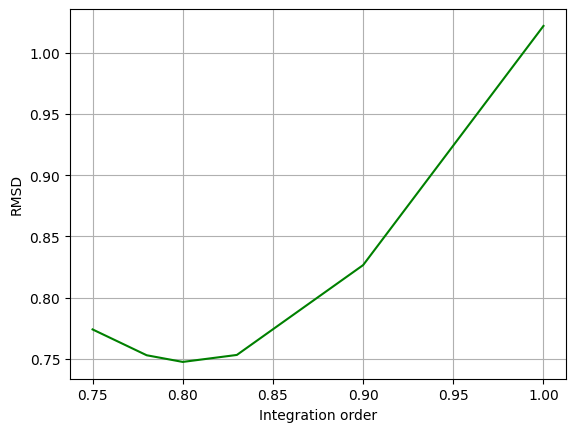}
        \caption{}
    \end{subfigure}
    \begin{subfigure}{0.48\textwidth}
        \includegraphics[width=\linewidth]{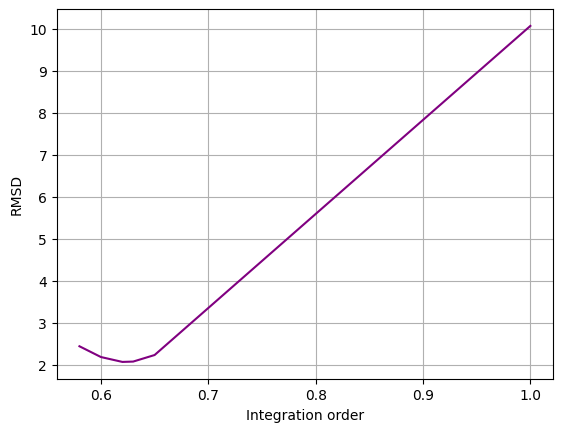}
        \caption{}
    \end{subfigure}
    \begin{subfigure}{0.48\textwidth}
        \includegraphics[width=\linewidth]{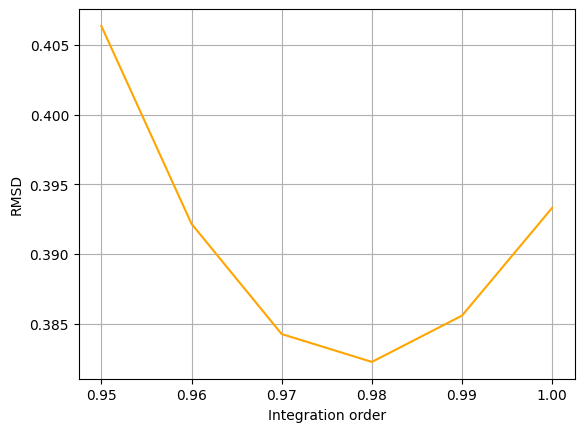}
        \caption{}
    \end{subfigure}
    \begin{subfigure}{0.48\textwidth}
        \includegraphics[width=\linewidth]{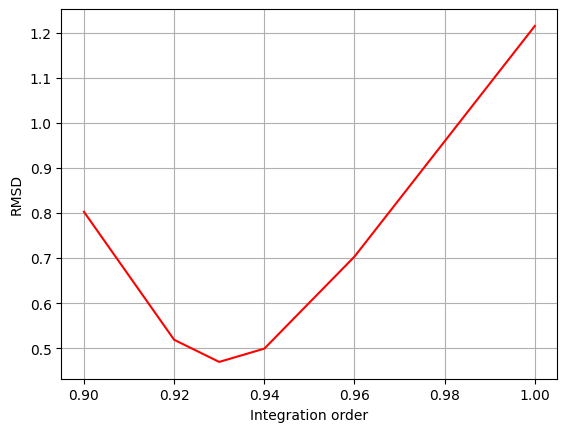}
        \caption{}
    \end{subfigure}

    \caption{RMSD evolution for different fractional orders in the case of the: (a) patient one; (b) patient two; (c) patient three; (d) patient four; (e) patient five}
    \label{fig:rmsd}
\end{figure}

For each fractional order analyzed, the corresponding RMSD value was computed by comparing the numerical solution of the model with the experimental tumor volume data. The impact of the fractional order on the quality of the fit is illustrated in Figure~\ref{fig:rmsd}, where it can be observed that an optimal fractional order exists for each patient, corresponding to the minimum RMSD value. This behavior indicates that the fractional-order model provides a better representation of the tumor growth dynamics than the classical integer-order model, highlighting the relevance of fractional dynamics in capturing patient-specific growth patterns.

Additionally, Table~\ref{tab:three} presents the fractional order $\alpha$ and the corresponding RMSD values for the six best approximations to the experimental tumor volume data. It can be observed that, for all patients, the fractional-order model yields lower RMSD values compared to the integer-order case ($\alpha = 1$), indicating an improved agreement with the experimental tumor growth dynamics.

For Patient~1, the RMSD is reduced by $69.23\%$ when using the optimal fractional order $\alpha = 0.94$. For Patient~2, a reduction of $26.81\%$ is achieved at the optimal fractional order $\alpha = 0.80$. In the case of Patient~3, a pronounced improvement is observed, with the RMSD decreasing by $79.48\%$ for $\alpha = 0.62$. Patient~4 shows a comparatively modest enhancement, with an RMSD reduction of $2.8\%$ at $\alpha = 0.98$. Finally, for Patient~5, the RMSD is reduced by $61.31\%$ at the optimal fractional order $\alpha = 0.93$.

The nonuniform magnitude of these improvements reflects the intrinsic heterogeneity of tumor growth dynamics and individual treatment responses. The variability in the optimal fractional order across patients indicates that the memory effects introduced by the fractional derivative play a patient-specific role in shaping tumor evolution. Patients exhibiting larger RMSD reductions appear to benefit more strongly from nonlocal temporal effects, while smaller improvements suggest dynamics closer to the classical integer-order regime. Overall, these findings reinforce the suitability of fractional-order models as a flexible and powerful framework for capturing individualized tumor growth patterns.

\begin{table}[H]
\centering
\begin{tabular}{|cc||cc||cc||cc||cc|}
\hline
\multicolumn{10}{|c|}{\textbf{Patient}} \\
\cline{1-10}
\multicolumn{2}{|c||}{\textbf{1}} 
 & \multicolumn{2}{|c||}{\textbf{2}} 
 & \multicolumn{2}{|c||}{\textbf{3}} 
 & \multicolumn{2}{|c||}{\textbf{4}} 
 & \multicolumn{2}{|c|}{\textbf{5}} \\
\cline{1-10}
$\boldsymbol{\alpha}$ & \textbf{RMSD}
& $\boldsymbol{\alpha}$ & \textbf{RMSD}
& $\boldsymbol{\alpha}$ & \textbf{RMSD}
& $\boldsymbol{\alpha}$ & \textbf{RMSD}
& $\boldsymbol{\alpha}$ & \textbf{RMSD} \\
\hline
1.00 & 0.013
& 1.00 & 1.022
& 1.00 & 10.062
& 1.00 & 0.393
& 1.00 & 1.215 \\
0.95 & 0.005
& 0.90 & 0.827
& 0.65 & 2.228
& 0.99 & 0.386
& 0.96 & 0.703 \\
\textbf{0.94} & \textbf{0.004}
& 0.83 & 0.753
& 0.63 & 2.072
& \textbf{0.98} & \textbf{0.382}
& 0.94 & 0.499 \\
0.93 & 0.005
& \textbf{0.80} & \textbf{0.748}
& \textbf{0.62} & \textbf{2.064}
& 0.97 & 0.384
& \textbf{0.93} & \textbf{0.470} \\
0.92 & 0.006
& 0.78 & 0.753
& 0.60 & 2.181
& 0.96 & 0.392
& 0.92 & 0.519 \\
0.90 & 0.009
& 0.75 & 0.774
& 0.58 & 2.438
& 0.95 & 0.406
& 0.9 & 0.802 \\
\hline
\end{tabular}
\caption{RMSD values for different fractional orders $\alpha$ for five patients. The minimum RMSD for each patient is highlighted in bold.}
\label{tab:three}
\end{table}

\begin{figure}[H]
    \centering
    
    \begin{subfigure}{0.48\textwidth}
        \includegraphics[width=\linewidth]{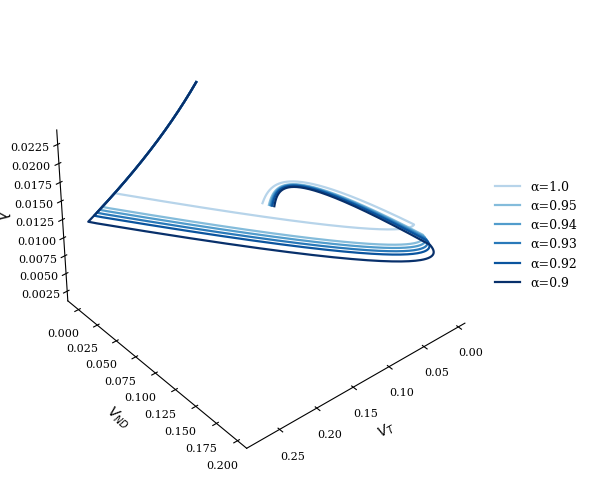}
        \caption{}
    \end{subfigure}
    \begin{subfigure}{0.48\textwidth}
        \includegraphics[width=\linewidth]{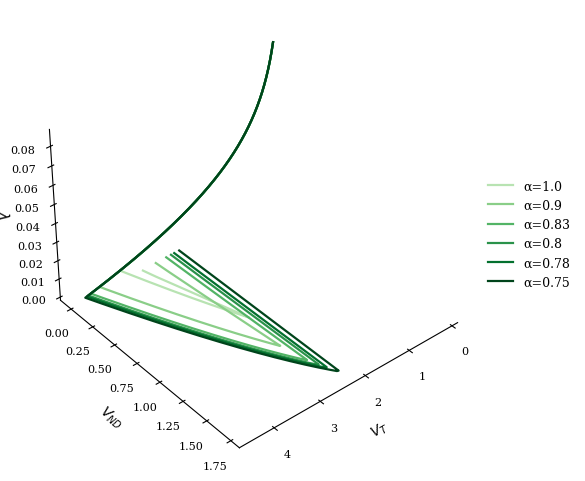}
        \caption{}
    \end{subfigure}
    \begin{subfigure}{0.48\textwidth}
        \includegraphics[width=\linewidth]{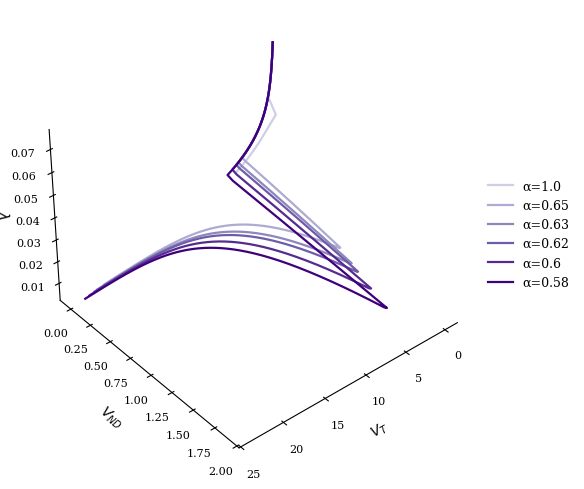}
        \caption{}
    \end{subfigure}
    \begin{subfigure}{0.48\textwidth}
        \includegraphics[width=\linewidth]{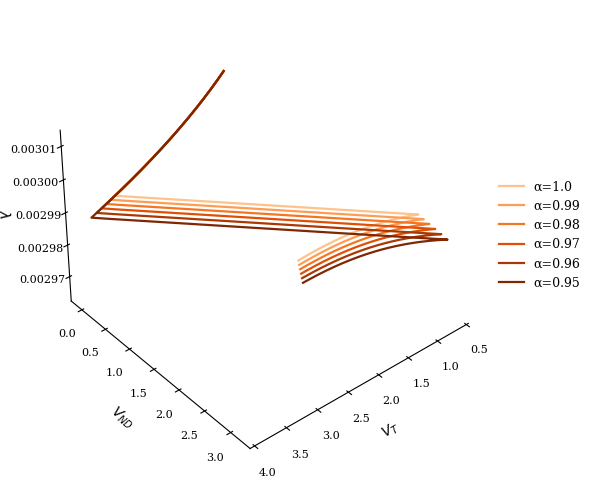}
        \caption{}
    \end{subfigure}
    \begin{subfigure}{0.48\textwidth}
        \includegraphics[width=\linewidth]{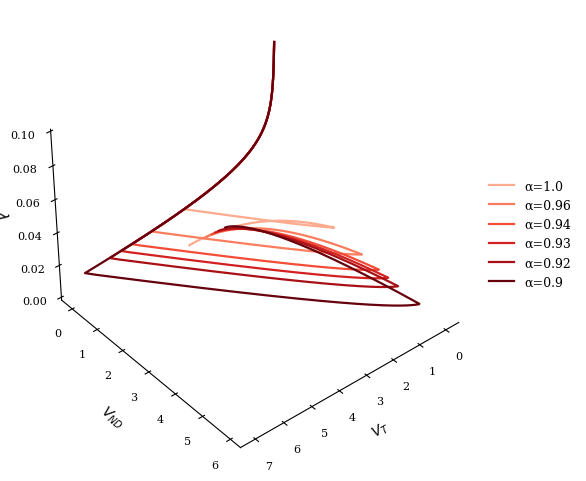}
        \caption{}
    \end{subfigure}

    \caption{Phase--space portraits for: (a) patient one; (b) patient two; (c) patient three; (d) patient four; (e) patient five}
    \label{fig:resultados_espacio}
\end{figure}

Figure~\ref{fig:resultados_espacio} presents the phase--space portraits obtained from the numerical simulations for the five patients, corresponding to different values of the fractional order $\alpha$. Each panel illustrates the joint evolution of the tumor volume and the associated state variables, allowing a geometric interpretation of the tumor dynamics under fractional-order modeling.

A clear dependence of the phase--space structure on the fractional order $\alpha$ is observed in all cases. As $\alpha$ decreases from the classical integer-order case ($\alpha = 1$), the trajectories exhibit noticeable changes in curvature, contraction, and transient behavior. 

The phase-space trajectories  reveal that tumor growth dynamics are deeply sensitive to the fractional order $\alpha$. As $\alpha$ decreases from the classical case to more memory-persistent states, distinct patterns emerge reflecting the intrinsic heterogeneity of the biological system.

Notably, the integer-order case typically exhibits sharper trajectories and more abrupt changes in phase--space direction. In contrast, the fractional-order models generate smoother, more curvilinear evolutions. This behavior is a direct consequence of the \textit{nonlocal temporal effects} and the "memory" characteristic of the Caputo derivative. Biologically, this smoothness represents the \textit{viscoelastic nature of the tumor mass} and the delayed cellular responses to microenvironmental stress, such as hypoxia or vascular adaptation. For lower values of $\alpha$, the trajectories converge more slowly toward the stable manifold, displaying broader exploration. This suggests that fractional dynamics capture a "biological inertia" where the tumor's past states act as a buffer against rapid changes.

These findings reinforce the hypothesis that fractional-order models provide a more robust framework for patient-specific modeling, as they account for the accumulated effects of treatment and cellular history that integer-order models inherently overlook.

\begin{figure}[H]
    \centering
    
    \begin{subfigure}{0.32\textwidth}
        \includegraphics[width=\linewidth]{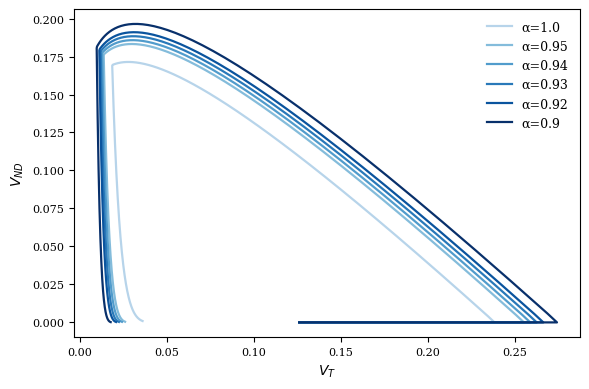}
        \caption{}
    \end{subfigure}
    \begin{subfigure}{0.32\textwidth}
        \includegraphics[width=\linewidth]{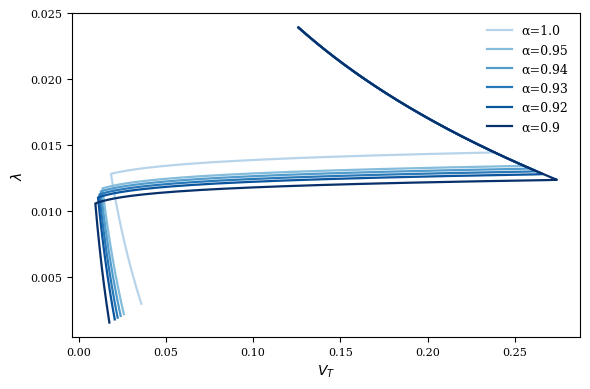}
        \caption{}
    \end{subfigure}
    \begin{subfigure}{0.32\textwidth}
        \includegraphics[width=\linewidth]{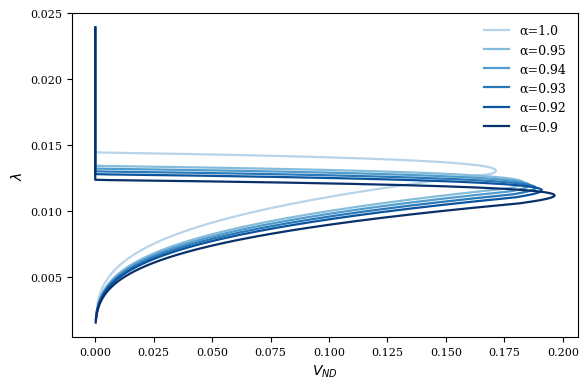}
        \caption{}
    \end{subfigure}

    \caption{Patient 1}
    \label{fig:resultados_patient_1}
\end{figure}

\begin{figure}[H]
    \centering
    
    \begin{subfigure}{0.32\textwidth}
        \includegraphics[width=\linewidth]{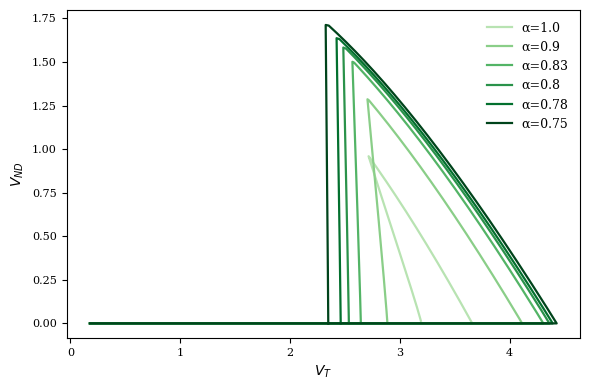}
        \caption{}
    \end{subfigure}
    \begin{subfigure}{0.32\textwidth}
        \includegraphics[width=\linewidth]{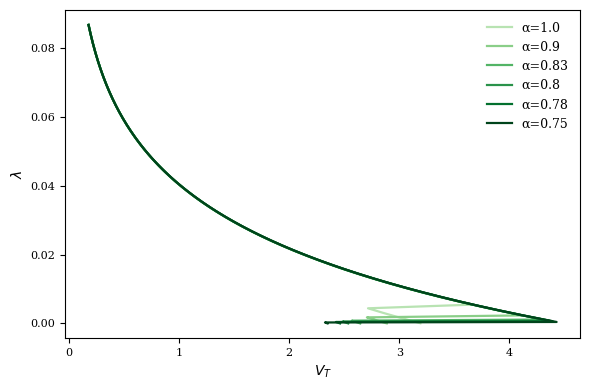}
        \caption{}
    \end{subfigure}
    \begin{subfigure}{0.32\textwidth}
        \includegraphics[width=\linewidth]{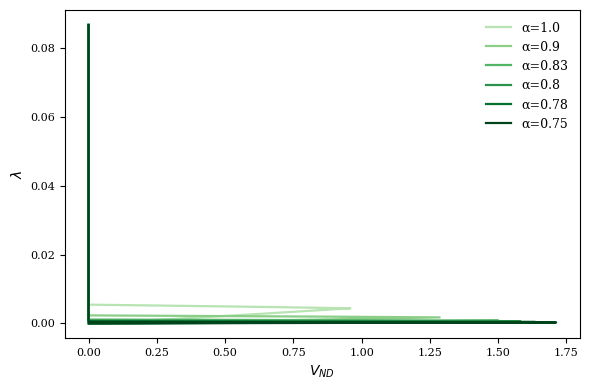}
        \caption{}
    \end{subfigure}

    \caption{Patient 2}
    \label{fig:resultados_patient_2}
\end{figure}

\begin{figure}[H]
    \centering
    
    \begin{subfigure}{0.32\textwidth}
        \includegraphics[width=\linewidth]{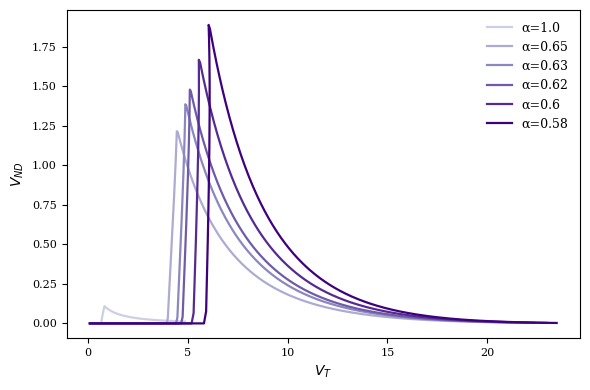}
        \caption{}
    \end{subfigure}
    \begin{subfigure}{0.32\textwidth}
        \includegraphics[width=\linewidth]{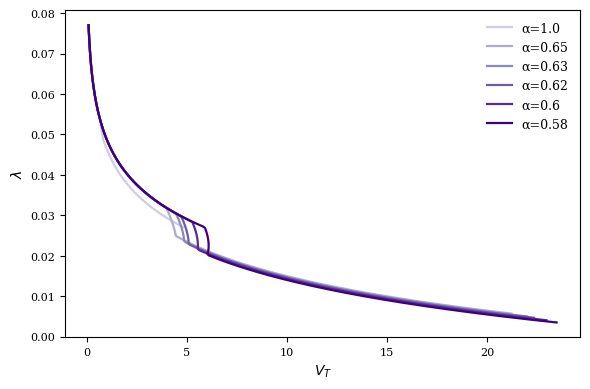}
        \caption{}
    \end{subfigure}
    \begin{subfigure}{0.32\textwidth}
        \includegraphics[width=\linewidth]{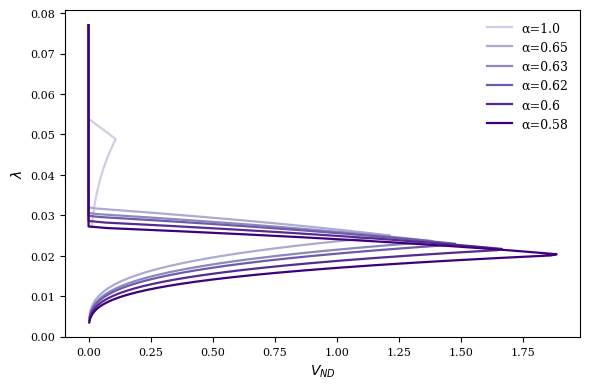}
        \caption{}
    \end{subfigure}

    \caption{Patient 3}
    \label{fig:resultados_patient_3}
\end{figure}

\begin{figure}[H]
    \centering
    
    \begin{subfigure}{0.32\textwidth}
        \includegraphics[width=\linewidth]{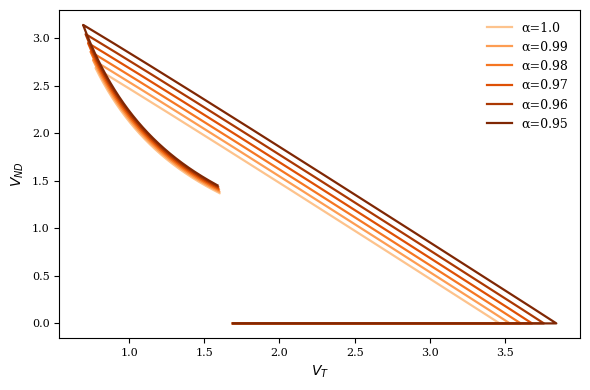}
        \caption{}
    \end{subfigure}
    \begin{subfigure}{0.32\textwidth}
        \includegraphics[width=\linewidth]{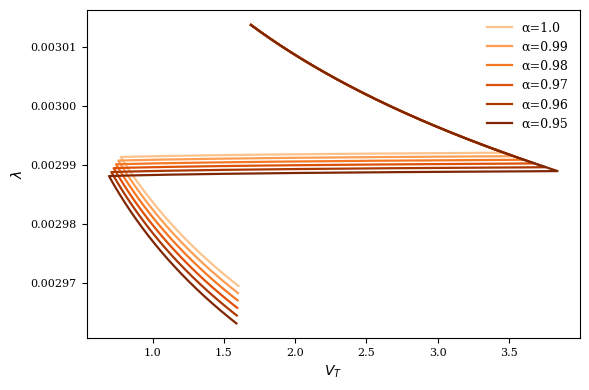}
        \caption{}
    \end{subfigure}
    \begin{subfigure}{0.32\textwidth}
        \includegraphics[width=\linewidth]{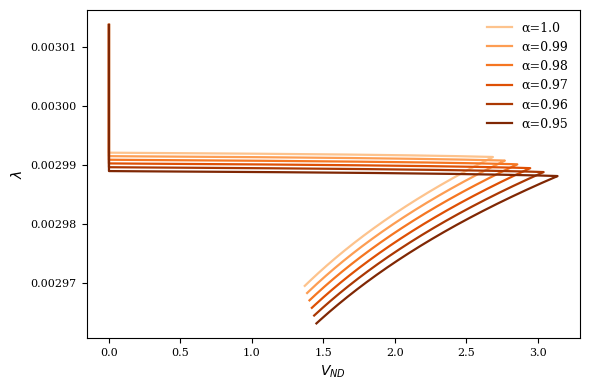}
        \caption{}
    \end{subfigure}

    \caption{Patient 4}
    \label{fig:resultados_patient_4}
\end{figure}

\begin{figure}[H]
    \centering
    
    \begin{subfigure}{0.32\textwidth}
        \includegraphics[width=\linewidth]{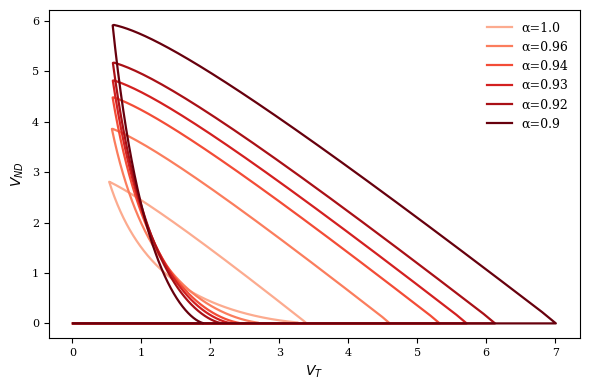}
        \caption{}
    \end{subfigure}
    \begin{subfigure}{0.32\textwidth}
        \includegraphics[width=\linewidth]{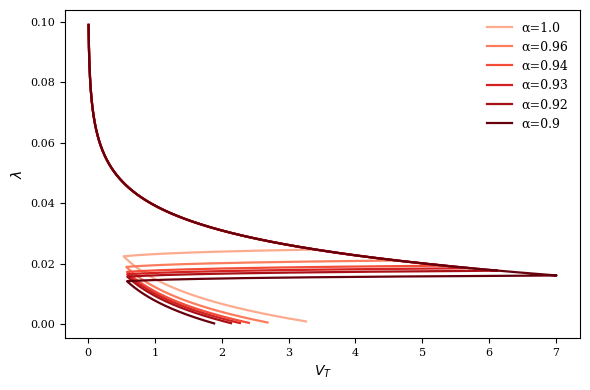}
        \caption{}
    \end{subfigure}
    \begin{subfigure}{0.32\textwidth}
        \includegraphics[width=\linewidth]{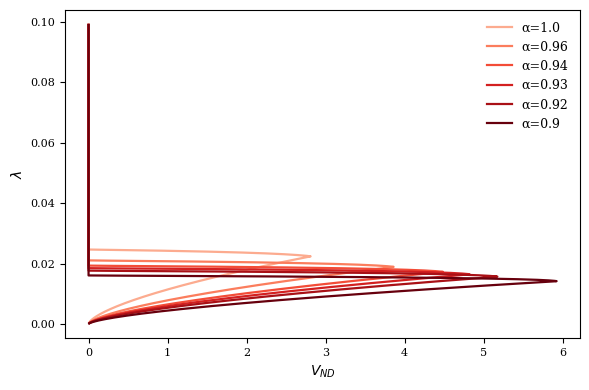}
        \caption{}
    \end{subfigure}

    \caption{Patient 5}
    \label{fig:resultados_patient_5}
\end{figure}

Figures~\ref{fig:resultados_patient_1} through~\ref{fig:resultados_patient_5} illustrate the phase--space projections for Patients~1–5 across various fractional orders $\alpha$. These visualizations reveal how the order of the derivative acts as a fundamental scaling parameter that reshapes the geometry of the system's attractors.

\begin{itemize}
    \item \textit{Projection (a): $V_T$ vs. $V_{ND}$.} This projection highlights the coupling between tumor volume and the nonproliferative population. We observe that as $\alpha$ decreases, there is a significant expansion in the spatial spread and curvature of the paths. While the integer-order case ($\alpha = 1$) follows a more constrained and direct manifold, fractional-order dynamics exhibit a "biological inertia" that forces the system to explore a wider phase space. This expansion reflects the increased complexity of cellular interactions in heterogeneous micro-environments, which classical models tend to oversimplify.
    
    \item \textit{Projection (b): $V_T$ vs. $\lambda$.} This plane emphasizes the temporal evolution of growth rates. A key finding here is the regularization effect: the integer-order case often displays sharp, abrupt transitions, whereas fractional models produce smoother, more gradual trajectories. This behavior is characteristic of \textit{nonlocal temporal effects}, where the current state of the tumor is moderated by its entire developmental history. 
    
    \item \textit{Projection (c): $V_{ND}$ vs. $\lambda$.} The nonproliferative population and growth rate relationship reveals a stabilizing effect inherent to fractional dynamics. Trajectories cluster more tightly as $\alpha$ deviates from unity, suggesting that the fractional derivative acts as a damping mechanism against sharp oscillations. Biologically, this represents the stabilizing influence of accumulated biological effects and delayed clearance processes, providing a more robust representation of long-term tumor stability.
\end{itemize}

In summary, the transition from integer-order to fractional-order dynamics represents a shift from classical diffusion to \textit{anomalous sub-diffusion} and memory-persistent growth. The clear geometric divergence between these models reinforces the necessity of fractional-order formulations; they do not merely adjust the rate of growth but fundamentally redefine the \textit{pathway} toward equilibrium. This patient-specific variation in phase-space exploration is consistent with our RMSD findings, confirming that the fractional order $\alpha$ serves as a critical biomarker for individualized tumor characterization and clinical trajectory prediction.

The phase--space analysis provides complementary qualitative evidence supporting the quantitative results obtained from the RMSD analysis. The observed geometric differences confirm that fractional-order modeling not only improves data fitting but also fundamentally alters the dynamical structure of tumor growth, offering a more flexible and realistic representation of tumor evolution across different patients.

\section{Conclusions}\label{sec:concl}

The comparison between numerical simulations and experimental tumor volume data demonstrates that fractional-order models consistently outperform their integer-order counterparts. For all patients analyzed, the Root Mean Square Deviation (RMSD) is significantly reduced when an optimal fractional order $\alpha<1$ is employed, highlighting the relevance of nonlocal temporal effects in capturing patient-specific tumor growth behavior. The variability observed in the optimal fractional order across patients reflects the intrinsic heterogeneity of tumor dynamics and treatment response.

Phase--space analyses further support these findings by revealing substantial geometric differences between integer-order and fractional-order trajectories. Notably, the integer-order case ($\alpha = 1.0$) typically exhibits \textit{sharper trajectories and more abrupt changes} in phase--space direction. This memoryless behavior reflects an instantaneous response to system dynamics, leading to the angular transitions.

In contrast, as the fractional order $\alpha$ decreases, the trajectories become \textit{progressively smoother and more coherent}. This regularizing effect is a direct manifestation of the nonlocal temporal characteristics introduced by the fractional derivative. By incorporating the influence of past states into the current evolution, fractional dynamics act as a biological "damping" mechanism that attenuates sudden transients and rapid oscillations. 

From a medical perspective, this smoothness suggests that tumor growth is not a series of disconnected events, but a continuous process governed by cumulative cellular history and vascular adaptation. The observed patient-dependent variations in these portraits reinforce the conclusion that fractional-order models capture the inherent viscoelasticity and memory of real-world tumor dynamics far more accurately than classical integer-order formulations.

Additionally, this study analyzed a fractional-order tumor growth model with memory effects and solved it numerically using a second-order fractional Runge--Kutta scheme. The proposed numerical method, based on a truncated fractional Taylor expansion, provides a stable and accurate approximation of the underlying fractional dynamics and enables a systematic investigation of the influence of the fractional order on tumor evolution.

Although the results are encouraging, the present study is based on a limited number of patient datasets. Extending the analysis to a larger cohort and incorporating more extensive experimental data would allow for a more robust statistical validation of the proposed fractional-order framework and a deeper assessment of its predictive capabilities. Such an extension would also facilitate the identification of systematic trends in the optimal fractional order and strengthen the potential applicability of the model to personalized tumor growth modeling and treatment planning.

\section*{Acknowledgements}
The authors gratefully acknowledge Yoichi Watanabe for kindly providing the original experimental tumor volume data analyzed in this study. His contribution was essential for the validation and assessment of the proposed fractional-order tumor growth model.

This work was carried out as part of the Master's thesis of the first author, who was supported by the Master's fellowship CVU 2045917, SECIHTI, Mexico

\end{document}